\documentclass[12pt]{article}
\usepackage{latexsym}
\usepackage{graphicx}
\usepackage{caption2}
\usepackage{psfrag}
\usepackage{amsmath}
\newcommand{\be}{\begin{equation}}
\newcommand{\ee}{\end{equation}}
\newcommand{\bea}{\begin{eqnarray}}
\newcommand{\eea}{\end{eqnarray}}

\newcommand{\nn}{\nonumber}
\newcommand{\lesssim}{ {\
\lower-1.2pt\vbox{\hbox{\rlap{$<$}\lower5pt\vbox{\hbox{$\sim$}}}}\ } }
\newcommand{\gtrsim}{ {\
\lower-1.2pt\vbox{\hbox{\rlap{$>$}\lower5pt\vbox{\hbox{$\sim$}}}}\ } }

\def\slash{\!\!\!\!/ \ }

\input epsf

\begin{document}

\begin{titlepage}

\begin{flushright}
\end{flushright}
\vspace*{1.5cm}
\begin{center}
{\Large \bf Kaon mixing and the charm mass}\\[2.0cm]

{\bf O. Cat\`{a}} and {\bf S. Peris }\\[1cm]

 Grup de F{\'\i}sica Te{\`o}rica and IFAE\\ Universitat
Aut{\`o}noma de Barcelona, 08193 Barcelona, Spain.\\[0.5cm]

\end{center}

\vspace*{1.0cm}

\begin{abstract}

We study contributions to the $\Delta S=2$ weak Chiral Lagrangian producing
$K^0-\overline{K}^0$ mixing which are not enhanced by the charm mass. For the real part,
these contributions turn out to be related to the box diagram with up quarks but, unlike
in perturbation theory, they do not vanish in the limit $m_u \rightarrow 0$. They
increase the leading contribution to the $K_L-K_S$ mass difference by $\sim 10 \%$. This
means that short distances amount to $(90\pm 15) \%$ of this mass difference. For the
imaginary part, we find a correction to the $\lambda^{*2}_c m_c^2$ term of $- 5\%$ from
the integration of charm, which is a small contribution to $\epsilon_K$. The calculation
is done in the large-$N_c$ limit and we show explicitly how to match short and long
distances.

\end{abstract}

\end{titlepage}

\section{Introduction.}

The Standard Model induces strangeness changing transitions by two units through the
famous box diagram connecting a $K^0$ to a $\overline{K}^0$ first considered in the
pioneering paper by Gaillard and Lee in 1974\cite{Gaillard74}. This diagram is depicted
in Fig. 1 where, as one can see, all quarks $u,c$ and $t$ are allowed to run in the box.
The result of this diagram is
\begin{eqnarray}\label{one}
    &&\mathcal{H}^{S=2}_{eff}\approx\frac{G_F^2}{4\pi^2}
    \ \overline{s}_L\gamma^{\mu}d_L(x)\
    \overline{s}_L\gamma_{\mu}d_L(x)\nn \\
 && \qquad \times \quad  \ c(\mu)\ \Bigg\{\eta_1 \lambda_c^2 m^2_c+ \eta_2
    \lambda_t^2 \left(m^2_{t}\right)_{eff}+ 2 \eta_3 \lambda_c \lambda_t m^2_c
    \log\frac{m_t^2}{m_c^2}\Bigg\}\ +\ \mathrm{h.c.}\quad ,
\end{eqnarray}
where we have defined $\psi_{L,R}=\frac{1\mp\gamma_5}{2}\psi, \lambda_i=V^{*}_{is}
V_{id}$ and $m_{u,c}$ stand for the corresponding quark masses\footnote{For the meaning
of $c(\mu)$ and $\eta_i$, see below.}. In the case of the top we have defined the
effective mass\cite{Buras2},
\begin{equation}\label{two}
\left(m^2_t\right)_{eff}= 2.39 \ M_W^2 \left[\frac{m_t}{167\ GeV}\right]^{1.52},
\end{equation}
which takes into account that the top mass is heavier than the $W$. In the (unrealistic)
limit $M_W\gg m_t\gg m_c$ one would have $(m^2_t)_{eff}\rightarrow m_t^2$. Physical
observables such as the $K_L-K_S$ mass difference and $\epsilon_K$ get a contribution
from the real and imaginary part, respectively, of the matrix element
$<K^0|\mathcal{H}^{S=2}_{eff}|\overline{K}^0>$.

\begin{figure}
\renewcommand{\captionfont}{\small \it}
\renewcommand{\captionlabelfont}{\small \it}
\centering \psfrag{A}{$\overline{s}_L$}\psfrag{B}{$d_L$}\psfrag{C}{$d_L$}
\psfrag{D}{$\overline{s}_L$}\psfrag{E}{$u,c,t$}
\psfrag{F}{$u,c,t$}\psfrag{G}{$W$}\psfrag{H}{$W$}
\includegraphics[width=2.5in]{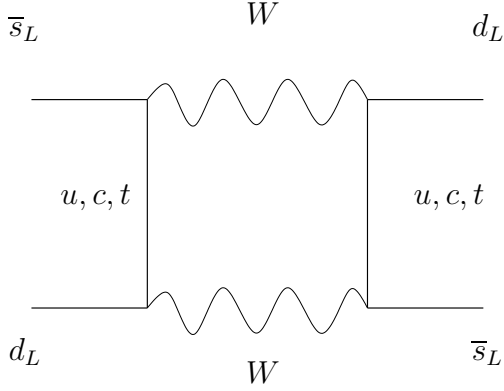}
\caption{Box diagram giving rise to $\Delta S=2$ transitions in the Standard
Model.}\label{plot1}
\end{figure}

As it is usually done, in Eq. (\ref{one}) we have neglected the up quark contribution to
the box diagram because it comes with a $m_u^2$ factor out front. Therefore, one could
naively conclude that the matrix element $<K^0|\mathcal{H}^{S=2}_{eff}|\overline{K}^0>$
is given by the operator (\ref{one}) up to very small corrections
$\mathcal{O}(m_u^2/m_c^2)$.

However, this argument is flawed because the above box-diagram was obtained assuming that
strong interactions are purely perturbative, which is obviously not true. Strong
interactions bind light quarks very tightly and produce a mass gap between the
$(\pi,K,\eta)$ octet and and the first hadronic resonance. Consequently, relative to the
kaon mass, there are two kinds of contributions to the $K^0-\overline{K}^0$ transition
depending on the hadronic scale governing the intermediate states. On the one hand, there
is the contribution in which the intermediate states are members of the Goldstone octet.
This contribution is given by the effect of the $\Delta S=1$ chiral Lagrangian acting
twice and is nonlocal at the scale of the kaon mass. In a chiral expansion, the leading
term in this contribution is of $\mathcal{O}(p^2)$ but vanishes due to the
Gell-Mann-Okubo mass relation. The remaining subleading term at $\mathcal{O}(p^4)$,
regretfully, depends on unknown low-energy coupling constants. An estimate of this term
was done in Ref. \cite{Donoghue} but the result depends on a cutoff and the uncertainties
are very large. It is clear, therefore, that a reliable calculation of these $(\Delta
S=1) \times (\Delta S=1)$ contributions remains a very important issue. In this work,
however, we shall not deal with it.

 On the other hand, there is the contribution in which all the intermediate states
are (much) heavier than the kaon. Among these there are of course the $c,b,t$ quarks and
the $W$ boson, but \emph{also} all the hadronic resonances with $u,d,s$ quark content.
This second type of contributions to the $K^0-\overline{K}^0$ transition gives rise to a
$\Delta S=2$ chiral Lagrangian. The above box diagram with an up quark is a
misrepresentation for the contribution coming from these resonances.

In the case of the contributions from $c,b,t$ one can of course include gluon corrections
in a perturbative way, and this has been done in Refs. \cite{Gilman-Wise,Nierste,Jamin}.
These perturbative corrections can be lumped into the $c(\mu)$ and $ \eta_i$
coefficients, with the result\cite{Buras2}

\begin{eqnarray}\label{three}
&&\eta_1=(1.32\pm 0.32)\left[\frac{1.3 GeV}{m_c(m_c)}\right]^{1.1}\quad ,\quad
\eta_2=0.57\pm 0.01\quad,\quad \eta_3=0.47\pm 0.05\  \nn\\
&&c(\mu)= \left(\alpha_s(\mu)\right)^{-\frac{9}{11N_c}}\ \left[1+
\frac{\alpha_s(\mu)}{\pi}\left( \frac{1433}{1936}+ \frac{1}{8}\ \kappa \right)  \right]\
,
\end{eqnarray}
where $\kappa=0\ (-4)$ in the naive dimensional regularization (resp. 't Hooft-Veltman)
schemes.

Obviously, in the case of the up quark contribution in Eq. (\ref{one}) there is no point
in including any perturbative gluon correction to it. Furthermore, since the up quark
contribution is proportional to $\lambda_u^2$, which is real, it can only contribute to
the $K_L-K_S$ mass difference. Then, the following natural question arises: since $m_u^2$
is misleading, what is the true size of this up quark contribution to the $K_L-K_S$ mass
difference? As we have argued above, there is a nonperturbative contribution which comes
from integrating out hadronic resonances. Therefore, in principle, there could be a
contribution proportional to a typical hadronic scale $\Lambda_{QCD}\sim 1$ GeV, so that
the corrections to the $K_L-K_S$ mass difference would become
$\mathcal{O}(\Lambda_{QCD}^2/m_c^2)$ and could be very important.

Furthermore, since $\lambda_{c,t}$ have a large imaginary part, there is an important
contribution to the $\epsilon_K$ parameter in the terms proportional to $m_c^2$ in Eq.
(\ref{one}). Could hadronic effects proportional  to $\Lambda_{QCD}^2$ produce large
corrections to, e.g., the imaginary part of the $\lambda_c^2 m_c^2$ term in (\ref{one})?

As a matter of fact, both questions above are related because they require considering
the problem of subleading corrections to the large-$m_c$ expansion. Somewhat
surprisingly, not much work has been devoted to the study of $1/m_c$ corrections. Apart
from the problem of $K^0-\overline{K}^0$ mixing, we are only aware of Refs. \cite{Falk,
Buchalla,Pivovarov2,Gerard} where these corrections have been considered. In the specific
case of the $K^0-\overline{K}^0$ matrix element, Bijnens, Gerard and Klein made an
estimate in Ref. \cite{Bijnens-Gerard}. They restricted themselves to the contribution
coming solely from Goldstone loops, selecting the leading topologies in the large-$N_c$
limit, and using a sharp cutoff to regulate the ultraviolet divergences which resulted.
In this way they obtained  a positive correction to the quadratic charm mass contribution
to the $K_L-K_S$ mass difference of the form $\lambda_c^2 (m_c^2 + \Lambda_H^2)$, with
$\Lambda_H \sim 0.5$ GeV. Although there was no matching between short and long distances
and, as a consequence, the physical result for the $K^0-\overline{K}^0$ matrix element
was cutoff dependent, there was reasonable numerical stability against variations of the
cutoff in the neighborhood of this value for $\Lambda_H $.

Pivovarov \cite{Pivovarov} considered the OPE and obtained matching at the level of quark
diagrams (in $\overline{MS}$) between the contribution to the box diagram with
intermediate charm and the contribution with only up quarks. However his calculation did
not show how to implement matching with mesons and, in particular, did not include the
constraints coming from chiral symmetry in the low-energy region. As a consequence, his
result was very sensitive to an ad-hoc infrared cutoff. Relative to the $\lambda_c^2
m_c^2$ contribution to the box diagram, he obtained a correction in the range $- 0.4
\lesssim \pm \frac{\Lambda_H^2}{m_c^2} \lesssim +0.1$, depending on the choice for this
infrared regulator.

It seems to us, therefore, that understanding the size of these corrections is an
interesting issue. In this work we shall reconsider the calculations of Refs.
\cite{Bijnens-Gerard, Pivovarov} in a framework which incorporates the correct infrared
and ultraviolet behavior and which allows matching between long and short distances. As
we shall see, the crucial ingredient for obtaining this matching is the consideration of
the constraints imposed by the Operator Product Expansion (OPE).

Let us back up a bit and describe how to organize the calculation using Effective Field
Theory. In an Effective Field Theory calculation one integrates out all fields whose
masses are larger than the kaon mass. This means that one has a theory in which the
quarks $c,b,t$, the gauge bosons $W,Z$ and the Higgs are no longer explicit degrees of
freedom in the Lagrangian. Only the $u,d,s$ quarks and the gluons propagate. A set of
matching conditions ensures that, even though the ultraviolet properties of the two
Lagrangians are very different, the physics from the Effective Lagrangian does not differ
from the one of the original Standard Model. The advantage of this way of organizing the
calculation is that one disentagles high- from low-energy scales, one at a time, and can
treat first all the heavy particles in a perturbative way, order by order in powers of
the coupling $\alpha_s$. This perturbative running from the $W$ scale down to the charm
scale is very well understood and under very good theoretical control thanks, among
others, to the work of Refs. \cite{Gilman-Wise,Nierste,Buras1,Jamin}.

After the integration of charm one still has a Lagrangian in terms
of the \emph{quarks} $u,d,s$, not the associated mesons, so that
the calculation of matrix elements between kaon states is still a
nontrivial task. However, kaons are pseudo-Goldstone bosons and
therefore their interactions can be organized in a Chiral
Lagrangian in powers of derivatives and masses. This means that
the quark operators obtained in the previous perturbative running
can be bosonized in terms of kaon fields as explicit degrees of
freedom. Once this is done, the calculation of a kaon matrix
element is straightforward. But in the construction of this chiral
Lagrangian, exactly as before, one has to make sure that a set of
matching conditions are satisfied. This point has been recognized
only recently\cite{Me, Eduardo}. These conditions are no longer
perturbative in powers of $\alpha_s$ and require a nonperturbative
treatment; with the help of the OPE, they control the interplay
between long and short distances.

Obviously these matching conditions cannot be solved exactly and one is forced to do some
approximations. Since, as usual, these conditions equate a Green's function computed in
terms of quarks to the same Green's function written in terms of mesons, it is very
convenient to use an approximation which can be carried out for both degrees of freedom.
The approximation we will use for this matching is the large-$N_c$ expansion in QCD
\cite{tHooft, Witten}.

The above program has been applied to other problems as well, including how to estimate
low-energy coupling constants in quenched QCD\cite{quench}. We refer to Ref.
\cite{theworks} for a list of references.

In section 2 we shall review a toy version of the calculation done by Gilman and Wise to
illustrate how the perturbative running is done, keeping only dimension-six operators.
Section 3 will be devoted to the inclusion in this perturbative running of
dimension-eight operators, suppressed by $1/m_c^2$ relative to the former, after the
integration of charm. Section 4 will contain the integration of resonances with the
corresponding matching to long distances. Finally, our conclusions will appear in section
5.

\begin{figure}
\renewcommand{\captionfont}{\small \it}
\renewcommand{\captionlabelfont}{\small \it}
\centering
\psfrag{M}{$\overline{s}_L$}\psfrag{N}{$d_L$}\psfrag{L}{$\overline{s}_L$}\psfrag{H}{$u,c$}
\psfrag{G}{$u,c$}\psfrag{O}{$d_L$}
\includegraphics[width=2.5in]{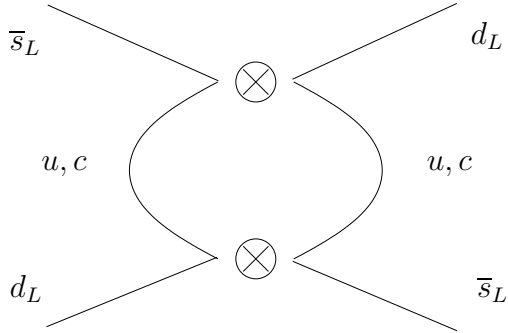}
\caption{Diagram contributing to the running of the operator (\ref{four}) in the
Effective Theory below $\mu_W \sim m_t \sim M_W$.}\label{plot2}
\end{figure}

\section{The Gilman-Wise calculation: simplified version.}

Let us imagine\footnote{This discussion parallels that of Ref. \cite{ManoharGeorgi}.}
 that the top were as heavy as the W, i.e. $m_t\sim M_W\sim \mu_W$. Then, at the
scale $\mu=\mu_W$, the box diagram in Fig. 1 would give rise to a local effective
operator given by
\begin{equation}\label{four}
    \mathcal{L}_{eff}=  \frac{G_F^2}{16\pi^2}
    \  c(\mu)\ \left\{\overline{s}\gamma^{\mu}(1-\gamma_5) d(x)\right\}^{2}\ .
\end{equation}
What this means is that shrinking the W propagator to a local Fermi interaction and
leaving out the top contribution differs from the full result in Fig. 1 by an amount that
can be encoded in the operator (\ref{four}), with the value of the coefficient given by
$c(m_t)= - m_t^2 \lambda_t^2$. This is the matching condition for the coefficient
$c(\mu)$ in the theory without top quarks and W bosons at $\mu=\mu_W$. This value of
$c(m_t)$ is to be taken as the boundary condition in the differential equations
controlling the running down to scales $\mu \ll \mu_W$.

In the region $m_c< \mu < \mu_W$, the running of the operator (\ref{four}) is given by
the $1/\varepsilon$ poles of the diagrams in Fig.~2 in dimensional regularization, where
only the quarks $c$  and $u$ run around the loop. This leads to the following
differential equation
\begin{equation}\label{five}
    \mu^{2}\frac{d}{d\mu^{2}}\ c(\mu)=-2 m_c^2 \lambda_c\left(\lambda_c+\lambda_u\right)
     -2 m_u^2 \lambda_u \left( \lambda_c +  \lambda_u\right)\ .
\end{equation}
Because in this calculation we are doing perturbation theory, the parameter $m_u$ is just
the up quark mass and its contribution may be safely neglected. With the above boundary
condition, this equation can be trivially integrated to (setting $m_u=0$)
\begin{equation}\label{six}
    c(\mu)=-m_t^2 \lambda_t^2 + 2 m_c^2 \lambda_c \left(\lambda_c+\lambda_u\right)
    \log\frac{m_t^2}{\mu^2}\ .
\end{equation}

At $\mu=m_c$, one has to integrate out the charm\footnote{In this simplified version the
quark bottom does not play any role.}. The diagram in Fig. 2 with only the c quark
running in the loop requires the coefficient in the operator (\ref{four}) in this new
theory \emph{without} charm to satisfy the matching condition $c(m_c)=-m_c^2
\lambda_c^2$, just like the condition from the integration of the top. Therefore, at
$\mu$ just below $m_c$ one has a theory with only the quarks $u,d,s$ and with the
coefficient of the operator (\ref{four}) given by
\begin{equation}\label{seven}
    c(\mu\lesssim m_c)=-m_t^2 \lambda_t^2 +2 m_c^2\lambda_c
    \left(\lambda_c+\lambda_u\right)\log\frac{m_t^2}{m_c^2}-m_c^2 \lambda_c^2\ .
\end{equation}
At $\mu< m_c$ the perturbative running of $c(\mu)$ is given by the diagram of Fig.~2 with
solely the up quark running in the loop. However, this contribution is proportional to
$m_u$. Setting $m_u=0$, the operator (\ref{four}) no longer runs. Therefore, the
coefficient $c(\mu)$ ``freezes out'' at a scale $\mu$ right below the charm mass, i.e. at
a scale which can still be considered perturbative, and it is given by Eq. (\ref{seven}).

Of course our calculation has been too simplistic in two respects. Firstly, $m_t$ is
larger than $M_W$ and, secondly, we neglected $\alpha_s$ corrections in the whole
discussion of running and matching. The modifications to amend this are nontrivial and
were done in Refs. \cite{Gilman-Wise,Nierste,Buras1,Jamin}. They give rise to the
coefficients $\eta_{1,2,3}$ and the effective mass $(m^2_t)_{eff}$ appearing in Eq.
(\ref{one}) (after use of the unitarity condition $\lambda_u+\lambda_c=-\lambda_t$ has
been made). Furthermore, $\alpha_s$ corrections ``defrost'' the coefficient $c(\mu)$
below $m_c$ and make it run in the region $M_{\rho}\lesssim \mu \lesssim m_c$ like in Eq.
(\ref{three}), where it matches to the weak chiral Lagrangian coming from long distances
\cite{Peris-deRafael}. However, and most importantly, all these short-distance
modifications are always proportional to the quark masses $m_t$ or $m_c$ and do not
change the conclusion of our simplified analysis: at this level there are no power
corrections of the scale $\Lambda_{QCD}\sim 1\ \mathrm{GeV}$. For these corrections to
show up, one needs to go to dimension-eight operators. This will be the subject of the
next section.

\section{Beyond Gilman-Wise: dimension-eight operators.}

In this section we shall consider the $\Delta S=2$ dimension-eight operators which are
generated by the successive integration of the top, the W, and finally charm.
Schematically, these operators are of the form
\begin{equation}\label{op1}
    \mathcal{L}^{\mathrm{dim\ 8}}_{\Delta S=2}\sim G_F^2\ \overline{s}_L\widehat{\mathrm{A}} d_L
    \ \overline{s}_L\widehat{\mathrm{B}} d_L\ ,
\end{equation}
with $(\widehat{\mathrm{A}};\widehat{\mathrm{B}})$ a gluon field strength $(G_{\mu\nu};
1)$ or two covariant derivatives $(D_{\mu};D_{\nu})$ acting on the quark fields.

In principle, the proliferation of scales and operators makes this a complicated problem.
To simplify matters, we shall neglect logarithmic corrections such as $\log (m_t/M_W)$ in
front of larger logarithms such as $\log (m_t/m_c)$. This means that we may integrate out
simultaneously the top and the W at a common scale, $\mu_W\sim m_t\sim M_W$.

At the scale $\mu_W$, $\Delta S=2$ dimension-eight operators may get generated from the
difference between the diagrams in Fig. 1 and their counterparts in the Effective Theory
without the top and the W boson. The different contributions can be broken down as
follows. Firstly, the top contribution will give terms proportional to $\lambda_t^2$; and
the mixed diagrams involving the top and the up and charm quarks will give terms
proportional to $\lambda_t (\lambda_c+\lambda_u)= - \lambda_t^2$, because of unitarity.
Notice that at the matching scale $\mu_W$ the $u,c$ quarks are effectively massless.
Moreover, the diagrams with only the $u,c$ quarks in the loop will be proportional to
$(\lambda_c+\lambda_u)^2=\lambda_t^2$. Therefore, the net result at $\mu_W$ is that all
contributions are modulated by the factor $\lambda_t^2 \sim \mathcal{O}(10^{-8})$ which
is tiny and can be neglected.  Notice that the $\lambda_t^2$ term in Eq. (\ref{one}) is
kept only because it is accompanied by the large $(m_t)^2_{eff}$ factor. This factor now
is impossible because, by dimensional analysis, there can be no $m_t$ in front of the
operator (\ref{op1}).

However, at the scale $\mu_W$ there are $\Delta S=1$ dimension-six operators which come
from $W$ exchange at $\mathcal{O}(G_F)$. The flavor structure of these $\Delta S=1$
operators is given schematically by
\begin{equation}\label{op2}
    \mathcal{L}^{\mathrm{dim\ 6}}_{\Delta S=1}\sim G_F
    \sum_{x,y=u,c}\ V_{xs}^{*} V_{yd}\ \overline{s}_L x_L\
    \overline{y}_L d_L\ ,
\end{equation}
with any combination of color indices.

 One should now run down to the scale $\mu=m_c$. In so doing, squaring one of the
operators (\ref{op2}) will produce a $\Delta S=2$ dimension-eight operator, such as
(\ref{op1}). However, unlike in the dimension-six case (\ref{five}), there can be no mass
factors in Eq. (\ref{op1}). Therefore, not only in the matching condition at $\mu_W$ but
also in the logarithmic running, the $c$ and $u$ quarks are effectively degenerate.
Furthermore, in the mixing of $\mathcal{L}^{\mathrm{dim\ 6}}_{\Delta S=1} \times
\mathcal{L}^{\mathrm{dim\ 6}}_{\Delta S=1} $ into the operator (\ref{op1}) the $x,y$
indices in (\ref{op2}) have to be contracted. Thus, this mixing arranges the GIM
cancelation in the sum
\begin{equation}\label{GIM}
   \left( \sum_{x=u,c} V_{xs}^{*} V_{xd}\right)^2\ ,
\end{equation}
and equals $(V_{ts}^{*} V_{td})^2=\lambda_t^2$, again. Since this is a consequence of
flavor symmetry, gluons cannot alter this conclusion.

At the scale $\mu=m_c$, charm gets integrated out. At this point we bring in the
large-$N_c$ approximation, which we shall use in the rest of this work. In the
large-$N_c$ limit, four-quark operators factorize and become a product of two independent
color singlets. This is true even in the presence of gluon operators. Furthermore, we
should keep in mind that we are eventually interested in matrix elements between a $K^0$
and a $\overline{K}^0$, to lowest order in the chiral expansion, i.e. $\mathcal{O}(p^2)$.
There are two ways in which we can get a dimension-eight operator: either as a product of
two dimension-four operators, or as a product of a dimension-3 operator times a
dimension-5 one. The first case yields contributions of higher chiral order, namely
$\mathcal{O}(p^4)$. This is because
\begin{equation}\label{current}
    \langle \overline{K}^0(p) \vert {\bar{s}}_L\gamma_{\nu}{\cal{D}}_{\mu}d_L\vert 0\rangle
     \sim \left( p_{\mu}p_{\nu}- \frac{p^2}{4} g_{\mu\nu}\right)\ ,
\end{equation}
as can be immediately seen by contracting with $g^{\mu\nu}$ and use of the equations of
motion in the chiral limit. So, only the possibility $\mathcal{O}(\mathrm{dim-3}) \times
\mathcal{O}(\mathrm{dim-5})$ is left. However, $\overline{s}
\widetilde{G}^{\mu\nu}\gamma_{\nu}d(x)$ is the \emph{only} dimension-five current
connecting a kaon to the vacuum\cite{Novikov}. Remarkably, this implies that there is
only one $\Delta S=2$ dimension-eight operator whose matrix element is of
$\mathcal{O}(p^2)$ between a $K^0$ and a $\overline{K}^0$ with momentum $p$. This
operator is given by
\begin{equation}\label{eight}
    \mathcal{L}_{eff}= - \frac{G_F^2}{3\pi^2}\ c_8(\mu)\mathcal{O}_8(\mu)+ \mathrm{h.c.}\quad
    \mathrm{where}\quad
    \mathcal{O}_8(x)= g_s \overline{s}_L \widetilde{G}^{\mu\nu}\gamma_{\nu}d_L(x)\
    \overline{s}_L\gamma_{\mu}d_L(x)\ ,
\end{equation}
where $g_s$ is the strong coupling constant. This operator (\ref{eight}) is CPS
symmetric, where CPS symmetry\cite{Bernard} is the combination of charge conjugation,
parity and the exchange of quarks $s\leftrightarrow d$ \footnote{We have found that the
operator basis used in Ref. \cite{Pivovarov} does not respect this CPS symmetry, the
offending operator being $\overline{s}_L \gamma_{\{\mu }D_{\nu \}} d_L \overline{s}_L
\gamma^{\{\mu}D^{\nu\}} d_L$, where $\{,\}$ means symmetrization. However, this operator
has vanishing matrix elements between a $K^0$ and a $\overline{K}^0$ in the large-$N_c$
limit and, therefore, drops out of our analysis.}.

\begin{figure}
\renewcommand{\captionfont}{\small \it}
\renewcommand{\captionlabelfont}{\small \it}
\centering \psfrag{M}{$\overline{s}_L$}\psfrag{G}{$u$}\psfrag{O}{$d_L$}
\psfrag{N}{$d_L$}\psfrag{H}{$u$}\psfrag{L}{$\overline{s}_L$}
\includegraphics[width=2.5in]{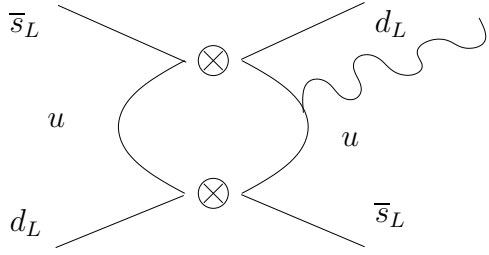}
\caption{Diagram responsible for the running of the operator (\ref{eight}) proportional
to $\lambda_u^2$.}\label{fig:plot}
\end{figure}

At the scale $\mu=m_c$, the diagram in Fig. 3 with charm and up quarks running in the
loop yields for the matching condition\footnote{The approximate relation
$\lambda_c^2+2\lambda_c\lambda_u\simeq -\lambda_u^2$ has been used.}

\begin{equation}\label{matchcond}
    c_8(m_c) = - \ \frac{7}{6}\ \lambda_u^2 - \frac{13}{6}\ \lambda_c \lambda_u .
\end{equation}
The term in Eq. (\ref{matchcond}) proportional to $\lambda_u^2$ is, of course, purely
real. However, the term proportional to $\lambda_c \lambda_u$ contains an imaginary part
which means that the parameter $\epsilon_K$ receives a contribution from the operator
$\mathcal{O}_{8}$ at the charm scale. Below $m_c$ only the up quark is active and,
therefore, the running of the operator $\mathcal{O}_8$ is proportional to $\lambda_u^2$,
and purely real. Consequently, we find that there are no further corrections to
$\epsilon_K$ coming from scales $\mu\sim  \Lambda_{QCD}\sim 1\ \mathrm{GeV}$ from the
integration of resonances.

Now, in the large-$N_c$ limit, gluon corrections and the contribution from the diagram in
Fig.~3 makes the operator (\ref{eight}) run according to (see Appendix A)
\begin{equation}\label{eleven}
    \mu^{2}\frac{d}{d\mu^2}\ c_8(\mu)=  \lambda_u^2 + \gamma_8 \alpha_s c_8(\mu)\ ,
\end{equation}
where
\begin{equation}\label{ten}
    \gamma_8 = \frac{4}{6 \pi}\ \left(N_c- \frac{1}{N_c}\right)
    \approx \frac{4}{6 \pi}N_c\ ,
\end{equation}
was first computed in Ref. \cite{Shuryak}. Now, one should integrate Eq. (\ref{eleven})
in the region $\mu \lesssim m_c$ subject to the boundary condition for $c_8(m_c)$ in Eq.
(\ref{matchcond}). To a sufficient approximation, one can actually neglect the $\alpha_s$
term and obtain the rather simple result
\begin{equation}\label{twelve}
    c_8(\mu)= -\ \frac{13}{6}\ \lambda_c \lambda_u +
    \lambda_u^2 \left(\log\frac{\mu^2}{m_c^2}- \frac{7}{6}\right)\ .
\end{equation}
Therefore, at scales $\mu \lesssim m_c$, the Effective Field Theory is given by
\begin{equation}\label{twelve-p}
    \mathcal{L}_{eff}=  - \frac{G_F^2}{3 \pi^2}\ c_8(\mu)\
    g_s\overline{s}_L \widetilde{G}^{\mu\nu}\gamma_{\nu}d_L\
    \overline{s}_L\gamma_{\mu}d_L
     - \frac{G_F}{\sqrt{2}} \lambda_u\  \overline{s}_L \gamma^{\mu}u_L\
    \overline{u}_L\gamma_{\mu}d_L + h.c.\ ,
\end{equation}
where the first (second) operator produces $\Delta S=2$ (respectively  $\Delta S=1$)
transitions. This second operator is none other than $Q_2$. In Eq. (\ref{twelve-p}) the
Wilson coefficient of $Q_2$, $z_2$, has been set to unity and  no other $\Delta S=1$
operator is considered. This is in accordance with the large-$N_c$ limit. In real life,
however, one has that $z_2(\mu\lesssim m_c)\simeq 1.3-1.4$\cite{Buras1}. Subleading
$1/N_c$ corrections are also naively expected to be of this size\footnote{This naive
expectation may turn out to be misleading when the leading term in $1/N_c$ is dominated
by a chiral scale such as $F_{\pi}$\cite{Hambye}. This is not the case here.}.

At scales $\mu\sim M_K \ll m_c$ perturbation theory ceases to be valid. One has to change
the description in terms of quarks and gluons in favor of meson fields as explicit
degrees of freedom. This will be the subject of the next section.

\section{Long-distance contribution.}

Below the charm mass the Effective Theory is given by the Lagrangian (\ref{twelve-p}) and
is written in terms of the quarks $u,d,s$ and gluons as explicit degrees of freedom.

In the large-$N_c$ limit, however, and in terms of mesons as degrees of freedom, this
Effective Theory contains, besides the octet of Goldstone bosons\footnote{In the
large-$N_c$ limit one should, strictly speaking, consider a nonet by including the
$\eta'$\cite{Witten2}.},  an infinite tower of resonances. In order to match to the
Lagrangian describing the only-Goldstone degrees of freedom, one must integrate out all
the resonance fields.

At the scale of the kaon, there is only one $\mathcal{O}(p^2)$  $\Delta S=2$ operator in
the weak chiral Lagrangian. It is given by
\begin{equation}\label{thirteen}
    \mathcal{L}_{eff}^{S=2}= \frac{G_F^2}{16 \pi^2}\ \ F_0^4
    \ \Lambda^2_{S=2} \ \mathrm{Tr}
    \left[\lambda_{32}\left(D^{\mu}U^{\dag}\right)U\lambda_{32}
    \left(D_{\mu}U^{\dag}\right)U\right] + \mathrm{h.c.}\
    ,
\end{equation}
where $[\lambda_{32}]_{ij}=\delta_{i3}\delta_{2j}$ is a (spurion) matrix in flavor space,
$F_0\simeq 0.087$ GeV is the pion decay constant (in the chiral limit) and $U$ is a
$3\times 3$ unitary matrix collecting the Goldstone boson degrees of freedom and
transforming as $U\rightarrow RUL^{\dag}$ under a flavor rotation $(R,L)$ of the group
$SU(3)_L\times SU(3)_R$. In Eq. (\ref{thirteen}) we have normalized with a dimensionful
coupling constant $\Lambda_{S=2}$. The bosonization of the dimension-six operator
obtained below the charm scale (section 2) yields for this coupling:
\begin{equation}\label{sd}
\Lambda_{S=2}^2|_{\mathrm{dim-6}}=\ g_{S=2}\ \Bigg[\eta_1 \lambda_c^2 m^2_c+ \eta_2
    \lambda_t^2 \left(m^2_{t}\right)_{eff}+ 2 \eta_3 \lambda_c \lambda_t m^2_c
    \log\frac{m_t^2}{m_c^2}\Bigg]\ ,
\end{equation}
where \begin{equation} g_{S=2}= 1+ \mathcal{O}\left(\frac{1}{N_c}\right)
\end{equation}
is dimensionless, and the $\mathcal{O}(1/N_c)$ corrections have been computed in Ref.
\cite{Peris-deRafael}.

 We shall now consider the contribution to $\Lambda^2_{S=2}$ due to the dimension-eight
operator obtained in the previous section. This is tantamount to matching the two
Lagrangians (\ref{twelve-p}) and (\ref{thirteen}). In order to do this it is convenient
to choose a Green's function  which does not contain any contribution from the singlet
combination $1_L\times 1_R$ under the flavor group $SU(3)_L\times SU(3)_R$ \cite{Hambye}.
This makes it explicit that the short distance properties be encoded in the Wilson
coefficients of the standard four-quark effective operators and makes the following OPE
analysis much simpler. As was done in Refs. \cite{Peris-deRafael} we select the Green's
function:
\begin{equation}\label{fourteen}
    \mathcal{G}^{S=2}_{\alpha\beta}(p)=\int d^4x\ e^{ipx}\
    \langle 0|\mathrm{T}\left\{R_{\alpha}^{ds}(x)R_{\beta}^{ds}(0)\right\} |0\rangle\ ,
\end{equation}
where $R_{\alpha}^{ds}(x)= \overline{d}_R \gamma_{\alpha} s_R(x)$.

The Effective Lagrangian (\ref{thirteen}) yields for the Green's  function
$\mathcal{G}^{S=2}_{\alpha\beta}(p)$ at small momentum $p$,
\begin{equation}\label{fifteen}
    \mathcal{G}^{S=2}_{\alpha\beta}(p)= - i \frac{G_F^2}{8 \pi^2}\ F_0^4\ \Lambda_{S=2}^2\
    \left(\frac{p_{\alpha}p_{\beta}}{p^2}-g_{\alpha\beta}  \right) + \mathcal{O}(p^2)\ .
\end{equation}
 In terms of the Lagrangian (\ref{twelve-p}), the function $\mathcal{G}_{\alpha\beta}^{S=2}(p)$ is given
by
\begin{eqnarray}\label{sixteen}
     \!\!\!\!\!\!\!\!\!\!\!\!\!\!\!\!\!\!\!
     \mathcal{G}^{S=2}_{\alpha\beta}(p)\!&=\!& - i\ \frac{G_F^2}{3 \pi^2}\ c_8(\mu) \int
     d^4xd^4z\ e^{ipx}\
     \langle0|\mathrm{T}\left\{R_{\alpha}^{ds}(x)
     \mathcal{O}_8(z)  R_{\beta}^{ds}(0)\right\}|0\rangle\nn\\
     &&\qquad \quad -i\ \frac{G_F^2}{2 \pi^2} \lambda_u^2 (4\pi \mu^2)^{\epsilon/2}
     \int dQ^2 (Q^2)^{1-\epsilon/2}
     \int d\Omega_q \Gamma_{\mu\nu\alpha}(q,p)\Gamma^{\nu\mu\beta}(q,p)
\end{eqnarray}
where an expansion for low momentum $p$ is understood, and only terms $\mathcal{O}(p^0)$
are retained. In Eq. (\ref{sixteen}) we have defined the following Green's function
\begin{equation}\label{seventeen}
    \Gamma_{\mu\nu\alpha}(q,p)= \lim_{p\rightarrow 0}\int d^4x d^4y\ e^{-i(qx+py)}
    \langle 0|\mathrm{\widehat{T}}\left\{R_{\alpha}^{ds}(y)L_{\mu}^{su}(x)L_{\nu}^{ud}(0)
    \right\}|0\rangle\ ,
\end{equation}
with $Q^2=-q^2$ being positive in the euclidean regime, $L_{\mu}^{su}(x)=\overline{s}_L
\gamma_{\mu} u_L(x)$, etc., and where $d\Omega_q$ stands for the solid angle integration
in $D=4-\epsilon$ dimensions, normalized so that
\begin{equation}\label{eightteen}
    \int d\Omega_q\ q_{\mu} q_{\nu}= \frac{q^2}{D}\ g_{\mu\nu}\ .
\end{equation}
The second term in (\ref{sixteen}) is the result of a double insertion of the operator
$Q_2$ (see Eq. (\ref{twelve-p})). In this second term an interesting subtlety arises.
Because of the triangle anomaly, defining the function $\Gamma_{\mu\nu\alpha}(q,p)$ in
Eq. (\ref{seventeen}) in terms of the ordinary covariant T product introduces
contributions which transform as $1_L\times 1_R$ under the flavor group $SU(3)_L\times
SU(3)_R$\footnote{After all, the anomaly can be computed in perturbation theory.}. Since
the Lagrangian we want to match to in Eq. (\ref{thirteen}) has an explicit $SU(3)_L\times
SU(3)_R$ flavor symmetry\footnote{Notice, in particular, that the result in Eq.
(\ref{fifteen}) is tranverse.}, it is clear that using a Green's function which violates
this symmetry because of the anomaly is an unnecessary complication one would like to
avoid. This is the reason why the Green's function $\Gamma_{\mu\nu\alpha}(q,p)$ in Eq.
(\ref{seventeen}) is defined in terms of a special T product, ``$\mathrm{\widehat{T}}$''.
This $\mathrm{\widehat{T}}$ product is the one which produces the factorized, or
left-right symmetric, form of the anomaly\cite{Bardeen} and secures that the Green's
function $\Gamma_{\mu\nu\alpha}(q,p)$ is actually anomaly free, satisfying naive Ward
identities\cite{KPPdeR04}(see Appendix B).

In the large-$N_c$ limit we can use factorization and rewrite the first term in Eq.
(\ref{sixteen}) containing the operator $\mathcal{O}_8$ as
\begin{equation}\label{18p}
    i\ \frac{G_F^2}{6 \pi^2}\ c_8(\mu)
    \left(\frac{p_{\alpha} p_{\beta}}{p^2}- g_{\alpha\beta}\right)\ F_0^4 \delta_K^2\ ,
\end{equation}
where, following Ref. \cite{Novikov}, we have defined a parameter $\delta_K^2$ as
\begin{equation}\label{delta}
    \langle0|g_s \overline{s}_L
    \widetilde{G}_{\mu\nu}^{a}\lambda_a\gamma^{\mu}d_L|K^0(p)\rangle= -i \sqrt{2}F_0
    \delta_K^2 p_{\nu}\ .
\end{equation}
In the equation above, $\lambda_a$ are the color Gell-Mann matrices, normalized so that
$\mathrm{Tr}\lambda_a \lambda_b=2 \delta_{ab}$. We shall next deal with the second term
in Eq. (\ref{sixteen}).

As we discuss in Appendix B, Ward identities restrict the form of the function
$\Gamma_{\mu\nu\alpha}(q,p)$ in the low-$p$ limit to be
\begin{eqnarray}\label{nineteen}
  \Gamma_{\mu\nu\alpha}(q,p) &=& \frac{F_0^2}{2 p^2 q^2}
  \left[p^2 q_{\mu} g_{\alpha\nu}+p^2 q_{\nu}g_{\alpha\mu}-p_{\alpha}p_{\beta}q_{\mu}
  -p_{\alpha} p_{\mu} q_{\nu}+ q_{\mu} q_{\nu}
  \left(\frac{p\cdot q }{q^2}p_{\alpha}-\frac{p^2}{q^2}q_{\alpha}\right)\right]\nn\\
   &+& I_1(Q^2) \left(q^2 g_{\mu\nu} -q_{\mu}q_{\nu}\right)
   \left(\frac{p\cdot q}{p^2}p_{\alpha}-q_{\alpha}\right)\nn\\
   &+& I_2(Q^2) \left[i \varepsilon_{\mu\nu\lambda\sigma}
   q^{\sigma} \frac{p_{\alpha}p^{\lambda}}{p^2}-
   i  \varepsilon_{\mu\nu\alpha\lambda} q^{\lambda}  \right]+ \mathcal{O}(p)\ ,
\end{eqnarray}
where $I_{1,2}(Q^2)$ are some unknown functions. The combination
appearing in Eq. (\ref{sixteen}), however, shows a much simpler
tensor structure, to wit
\begin{equation}\label{twenty}
    \int d\Omega_q \Gamma_{\mu\nu\alpha}(q,p) \Gamma^{\nu\mu\beta}(q,p)=
    \left(\frac{p_{\alpha} p_{\beta}}{p^2}-g_{\alpha\beta}\right)\ W(Q^2)+ \mathcal{O}(p^2)\ ,
\end{equation}
where
\begin{equation}\label{21}
    W(Q^2)= \frac{3}{4}\ I_1^2(Q^2)\ Q^6-
    \frac{3}{2}\ I_2^2(Q^2)\ Q^2
    + \frac{7}{16}\ \frac{F_0^4}{Q^2}\ .
\end{equation}

The integral over $Q^2$ appearing in the matching condition (\ref{sixteen}) requires
knowledge of the function $W(Q^2)$ for \emph{all} values of $Q^2$ and this is not
available. This is why computing electroweak matrix elements is difficult. However, we
know how the function $W(Q^2)$ behaves at low values of $Q^2$ because this is given by
Chiral Perturbation Theory. A straightforward, though somewhat lengthy, calculation leads
to (one can take $\epsilon=0$ here)
\begin{equation}\label{22}
    Q^2 W(Q^2) \approx  \frac{7}{16}F_0^4 -
    \left[\frac{3}{2} \left(\frac{N_c}{24\pi^2}\right)^2 -3
    L_9^2\right]Q^4+ \mathcal{O}(Q^6)\ ,
\end{equation}
where the first contribution in squared brackets is due to the Wess-Zumino term, with the
anomaly subtracted to conform with the definition of $\Gamma_{\mu\nu\alpha}(q,p)$ and the
$\mathrm{\widehat{T}}$ product (see Appendix B). The second term in (\ref{22}) stems from
the $L_9$ $\mathcal{O}(p^4)$ coupling in the Gasser-Leutwyler Lagrangian.

 Also, at large values of $Q^2$ the OPE governs an
expansion in powers of $1/Q^2$ and yields
\begin{equation}\label{23}
    Q^2 W(Q^2) \approx  \frac{F_0^4}{4}+ \frac{F_0^4}{3}\frac{\delta_K^2}{Q^2}
    \left(1+ \frac{\epsilon}{6} \right)+
    \mathcal{O}(\frac{1}{Q^4})\ .
\end{equation}

If we build an interpolator between the two regimes, our work is done. In this respect we
recall that, in the large-$N_c$ limit, the coupling $L_9$ does not run with the scale and
the function $W(Q^2)$ is a meromorphic function of the meson masses, i.e. it has poles at
the particles' masses, but no cut. For instance, Figure~4 shows the pole structure of the
function $\Gamma_{\mu\nu\alpha}$ in Eq. (\ref{seventeen}). Thus, according to the
Mittag-Leffler theorem\cite{Jeffreys}, the function $W(Q^2)$ can be written as an
infinite sum of the principal parts at all poles, i.e.
\begin{equation}\label{24}
    W(Q^2)=\sum_{i=1}^{\infty}\frac{r_i}{(Q^2+M^2_i)^{p_i}}\quad ,\quad p_i=1,2,3,...
\end{equation}

\begin{figure}
\renewcommand{\captionfont}{\small \it}
\renewcommand{\captionlabelfont}{\small \it}
\centering \psfrag{A}{$R_{ds}$}\psfrag{B}{$L$}
\psfrag{C}{$L$}\psfrag{D}{$R_{ds}$}\psfrag{F}{$L$}\psfrag{G}{$L$}
\includegraphics[width=1.5in]{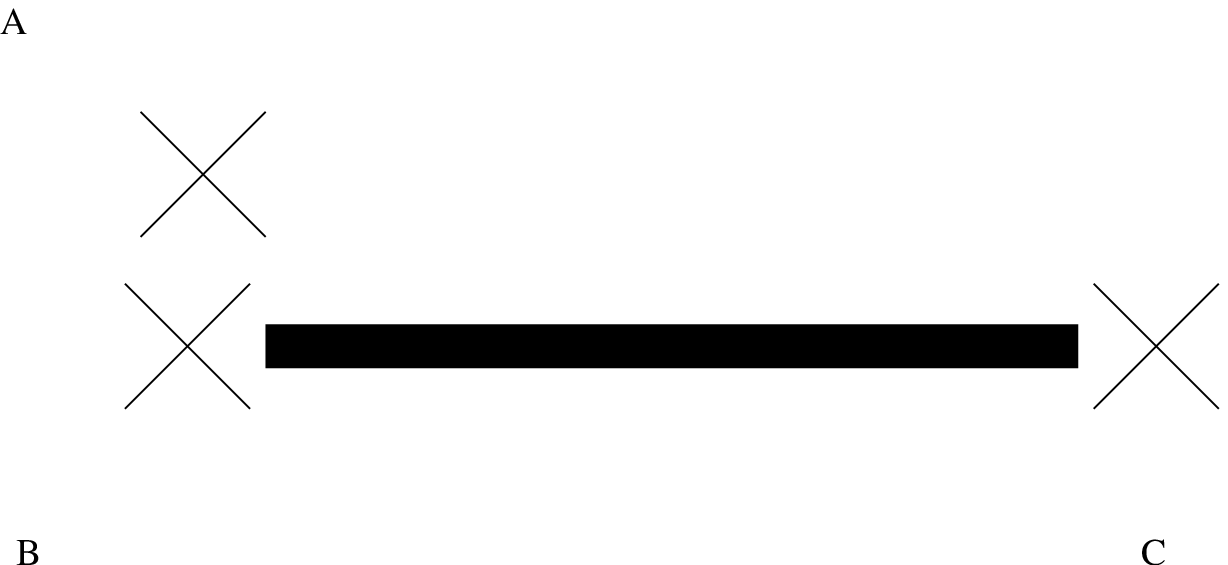}
\hspace{1cm}
\includegraphics[width=1.5in]{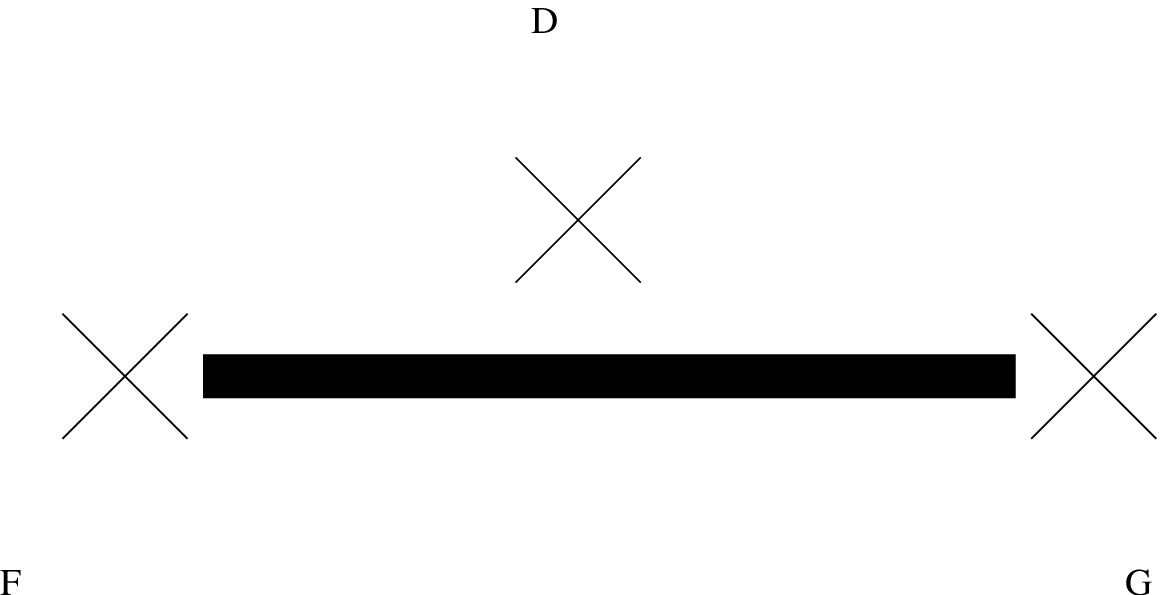}
\caption{Pole structure of the function $\Gamma_{\mu\nu\alpha}$ in Eq. (\ref{seventeen}).
``$L$'' stands for $L_{ud}$ or $L_{su}$.}\label{plot4}
\end{figure}

Lacking a solution to large-$N_c$ QCD, the value of the masses, $M_i$, and the residues,
$r_i$, remain unknown. However, using the constraints at low and large $Q^2$ in Eqs.
(\ref{22},\ref{23}) one may build a rational interpolator to the function $W(Q^2)$. This
interpolator consists in restricting the sum in Eq. (\ref{24}) to  a \emph{finite} sum,
adjusting the residues so that they match the two expansions given by
(\ref{22},\ref{23}). The rational interpolator so constructed constitutes an
approximation to large-$N_c$ QCD\cite{Me, Eduardo}. The simplest of such interpolators is
given by
\begin{equation}\label{25}
    Q^2 W(Q^2)_{HA}= a + \frac{A}{Q^2+M_V^2}+
    \frac{B}{(Q^2+M_V^2)^2}+\frac{C}{(Q^2+M_V^2)^3}+
    \frac{D}{(Q^2+M_V^2)^4}\ ,
\end{equation}
where we shall take $M_V\simeq 0.77$ GeV as an estimate of $M_{\rho}$ in the chiral and
large-$N_c$ limits. This way one may determine the 5 unknowns $a, A,B,C$ and $D$ from the
3 conditions from the chiral expansion (\ref{22}) and the 2 conditions from the OPE
(\ref{23}) as the solutions to the \emph{algebraic} equations:
\begin{eqnarray}\label{conditions}
  \frac{A}{M_V^2}+ \frac{B}{M_V^4}+\frac{C}{M_V^6}+\frac{D}{M_V^8} &=&
  \frac{7}{16} F_0^4 -a \nonumber \\
  \frac{A}{M_V^4}+ \frac{2B}{M_V^6}+\frac{3C}{M_V^8}+\frac{4D}{M_V^{10}} &=& 0\nonumber \\
  \frac{A}{M_V^6}+ \frac{3B}{M_V^8}+\frac{6C}{M_V^{10}}+\frac{10D}{M_V^{12}} &=& -\frac{3}{2}
  \ \frac{N_c^2}{(24 \pi^2)^2}+  3\ L_9^2 \nonumber\\
  a &=& \frac{F_0^4}{4} \nonumber\\
  A &=& \frac{F_0^4 \delta_K^2}{3} \left(1+ \frac{\epsilon}{6} \right)\ .
\end{eqnarray}

\begin{figure}
\renewcommand{\captionfont}{\small \it}
\renewcommand{\captionlabelfont}{\small \it}
\centering \psfrag{T}{\Large $Q^2$} \psfrag{Expressio}{\Large \!\!\!\!\!\!\!\!\!\!\!\!$\
\frac{16}{3}\frac{Q^2}{F_0^4}\ W_{HA}(Q^2)$} \vspace{1cm}
\includegraphics[width=4in]{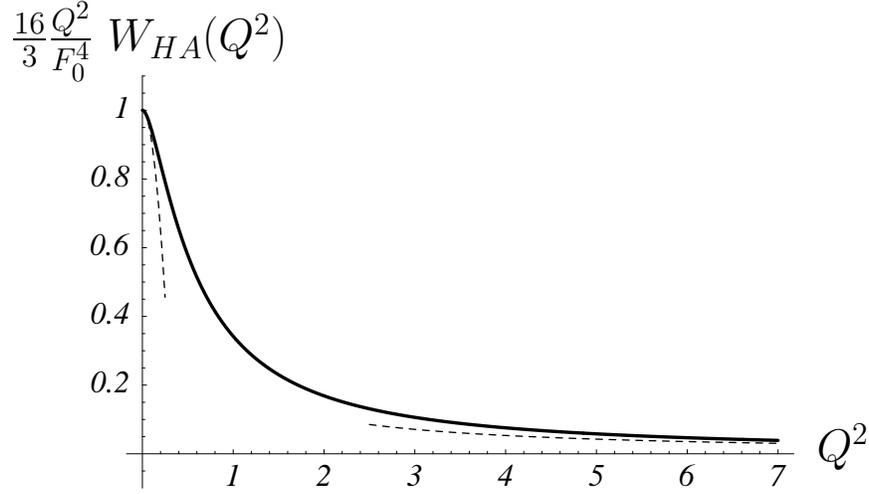}
\caption{Profile of the function $Q^2\ W_{HA}(Q^2)$ together with the chiral and OPE
behavior at low and high $Q^2$ (in $D=4$), respectively. The pion contribution has been
subtracted (the ``$a$'' term in Eq. (\ref{25})), according to the discussion in the
text.}\label{fig:plot}
\end{figure}

Notice that, since we use dimensional regularization, the pion contribution represented
by the $a$ term in Eq. (\ref{25}) gives a vanishing contribution to the integral
(\ref{sixteen}), although it does contribute to the chiral and OPE constraints in Eqs.
(\ref{conditions}). This is why the resulting function $Q^2 W_{HA}(Q^2)$, together with
the OPE and chiral extrapolations, are being plotted in Fig. 1 with the pion contribution
subtracted.

Equating  expressions (\ref{fifteen}) and (\ref{sixteen}), using (\ref{25}) in order to
approximate the integral over $Q^2$, one obtains the contribution from the effective
Lagrangian (\ref{twelve-p}) to the coupling constant $\Lambda^2_{S=2}$ in the weak chiral
Lagrangian (\ref{thirteen}) as
\begin{equation}\label{26}
    \Lambda^2_{S=2}|_{\mathrm{dim-8}}= \frac{4}{3}\delta_K^2 \left\{\lambda_u^2 \left( \log\frac{\mu^2}{M_V^2}  -
    \frac{1}{3}\right)- c_8(\mu)\right\}+ 4 \frac{\lambda_u^2}{F_0^4}
    \left[\frac{B}{M_V^2}+ \frac{C}{2 M_V^4} +
    \frac{D}{3 M_V^6} \right]
    \ .
\end{equation}
As one can see, the constraint of the OPE in the last equation in (\ref{conditions})
enforces the cancelation of the $\mu$ dependence from the Wilson coefficient $c_8(\mu)$,
Eq. (\ref{twelve}), with that coming from the integration of meson resonances, thereby
obtaining explicit matching between short and long distances.

The cancelation of the $\mu$ dependence agrees with that found in Ref. \cite{Pivovarov}
in terms of quark diagrams. However, Ref. \cite{Pivovarov} did not consider the
constraints imposed by the chiral behavior in Eq. (\ref{22}) and, as a consequence, the
result depended on an ad-hoc infrared regulator with the ensuing uncertainty in the
result.

Using $L_9= 7\times 10^{-3}$, $F_0=0.087\ \mathrm{GeV}$, $M_V=0.77\ \mathrm{GeV}$, as an
estimate of the vector resonance mass in the chiral and large-$N_c$ limits,
$\delta_K^2=0.12\pm 0.07\ \mathrm{GeV}^2$ \cite{delta}\footnote{Strictly speaking, the
operator on the left-hand side of (\ref{delta}) runs with the scale $\mu$. However, this
amounts to only a $10\%$ change if $\mu$ is varied in the range $1-2\ \mathrm{GeV}$, and
we shall neglect it.} and the charm mass $m_c(m_c)=1.3\ \mathrm{GeV}$,  we find that
\begin{equation}\label{result}
\Lambda^2_{S=2}|_{\mathrm{dim-8}}= \left\{\frac{26}{9}\ \lambda_c \lambda_u\  \delta_K^2
+ \lambda_u^2\ (0.95\ \mathrm{GeV})^2\right\}\left( 1 \pm 0.3\right)\ ,
\end{equation}
where the error is an estimate of the size of a typical $1/N_c$ correction\footnote{One
can make a naive check by building another interpolator imposing the extra condition of
having exact agreement with the OPE curve at, e.g., $Q^2=5\ \mathrm{GeV}^2$, which seems
to be large enough for the OPE to be trusted. This requires the introduction of an extra
pole which we took to be a single pole located at the A1 mass. The result obtained in
this case differed from the central value in (\ref{result}) by  $\sim 10\%$.}. Therefore,
as we discussed in the introduction, $\mathrm{Re}\ \Lambda^2_{S=2}|_{\mathrm{dim-8}} \sim
\lambda_u^2 \ \Lambda^2_{QCD}$ with $\Lambda_{QCD} \sim 1\ \mathrm{GeV}$, whereas
$\mathrm{Im}\ \Lambda^2_{S=2}|_{\mathrm{dim-8}} \sim \mathrm{Im}(\lambda_c \lambda_u)\
\delta_K^2$ with $\delta_K \sim 350\ \mathrm{MeV}$.

\section{Results.}

Armed with the Lagrangian (\ref{thirteen}), the dimension-six contribution (\ref{sd}),
and our result for the dimension-eight contribution (\ref{26}), one can now calculate the
matrix element $\langle K^0|\mathcal{H}_{eff}^{S=2}|\overline{K}^0\rangle$.

In the case of $\epsilon_K$, collecting all the terms, we find
\begin{eqnarray}\label{epsilo}
    \epsilon_K &=& e^{i\pi/4}\
    \frac{G_F^2 F_0^2 M_K}{6\sqrt{2}\pi^2 \left(M_{K_{L}}-M_{K_{S}}\right)}\ \times\\
    && \mathrm{Im}\ \left\{\frac{3}{4}\ g_{S=2} \Big[\eta_1 \lambda^{*2}_{c} m^2_c+ \eta_2
    \lambda^{*2}_t \left(m^2_{t}\right)_{eff}+ 2 \eta_3 \lambda^{*}_c \lambda^{*}_t m^2_c
    \log\frac{m_t^2}{m_c^2}\Big]+ \frac{3}{4}\ \Lambda^{*2}_{S=2}|_{dim-8}
    \right\}\ ,\nonumber
\end{eqnarray}
whereas for the Kaon mass difference, one has that
\begin{eqnarray}\label{mass}
    M_{K_{L}}-M_{K_{S}}&=&\frac{1}{M_K}\ \mathrm{Re}
    \langle K^0|\mathcal{H}_{eff}^{S=2}(0)|\overline{K}^0\rangle =
    \frac{G_F^2}{3\pi^2}F_0^2
    M_K \times \\
    &&\!\!\!\!\!\!\!\!\!\!\!\!\!\!\!\!\!\!\!\!\!\!\!\!\!\!\!\!\!\!\!\!\!\!\!\!
    \ \mathrm{Re}\ \left\{ \frac{3}{4}\ g_{S=2} \Big[\eta_1 \lambda_c^{*2} m^2_c
    + \eta_2
    \lambda_t^{*2} \left(m^2_{t}\right)_{eff}+ 2 \eta_3 \lambda_c^* \lambda_t^* m^2_c
    \log\frac{m_t^2}{m_c^2}\Big]+ \frac{3}{4}\ \Lambda^{*2}_{S=2}|_{dim-8}\right\}\ .
    \nonumber
\end{eqnarray}
In these two expressions $\Lambda^{2}_{S=2}|_{dim-8}$ is given by Eq. (\ref{result}). We
emphasize that these results are still in the chiral limit. We also remind the reader
that, in the large-$N_c$ limit, $g_{S=2}=1$. Nevertheless, some chiral and $1/N_c$
corrections to the contribution of dimension-six operators are known analytically. They
amount to the replacement $g_{S=2}\rightarrow \frac{4}{3}\widehat{B}^{\chi}_K
\frac{F_K^2}{F_0^2}$, where $\widehat{B}^{\chi}_K =0.36\pm 0.15$ \cite{Peris-deRafael}
and $F_K= 0.114\ \mathrm{GeV}$. Lattice calculations at the kaon mass modify the above
result by further changing $\widehat{B}^{\chi}_K \rightarrow \widehat{B}_K$, with
$\widehat{B}_K\simeq 0.86\pm 0.15$ \cite{Lellouch,Bijnens}.

After all these replacements are made, and feeding these expressions with $m_c=1.3\
\mathrm{GeV}, m_t=175\ \mathrm{GeV}, M_K=0.498\ \mathrm{GeV}$ and the central values of
the CKM matrix entries given by Ref. \cite{Buras2}, we find that the
$\Lambda^2_{S=2}|_{dim-8}$ contribution is a $ -5\%$ correction to the
$\eta_1\lambda^{*2}_{c} m_c^2$ term in $\epsilon_K$, which turns into a $+0.5 \%$
correction to $\epsilon_K$ itself. If the comparison was made in the strict chiral and
large-$N_c$ limits, the latter correction would become $+1\%$.

Concerning the kaon mass difference, we find that the $\Lambda^2_{S=2}|_{dim-8}$
contribution becomes a $+ 10 \%$ correction to the dimension-six result. This relative
correction becomes $ +25 \%$ in the strict chiral and large-$N_c$ limits.

We emphasize that the difference in size between the corrections to $\epsilon_K$ and the
kaon mass difference is mainly due to the fact that $\delta_K \sim 350 \ \mathrm{MeV}$
whereas $\Lambda_{QCD}\sim 1 \ \mathrm{GeV}$ is much larger (see Eq. (\ref{result})).

Using Eq. (\ref{mass}), we find $ M_{K_L}-M_{K_S}|_{\mathrm{SD}}= (3.1\pm 0.5)\times
10^{-15}\ \mathrm{GeV}$. Since, experimentally, we know that $M_{K_L}-M_{K_S}= (3.490\pm
0.006)\times 10^{-15}\ \mathrm{GeV}$\cite{RPP}, we conclude that short distances amount
to $(90 \pm 15)\%$ of this mass difference when the value $\widehat{B}_K\simeq 0.86\pm
0.15$ is used. The rest presumably must come from the double insertion of the $\Delta
S=1$ chiral Lagrangian at $\mathcal{O}(p^4)$, but this calculation is beyond the scope of
the present paper.

\vskip 2cm \textbf{Acknowledgements}

We thank M. Knecht and E. de Rafael for reading the manuscript and for their comments. We
particularly thank M. Knecht for asking us about the role played by the anomaly. This
work has been supported in part by TMR, EC-Contract No. HPRN-CT-2002-00311 (EURIDICE) and
by the research projects CICYT-FEDER-FPA2002-00748 and 2001-SGR00188.

\vskip 2cm

\appendix

\renewcommand{\theequation}{A.\arabic{equation}} \setcounter{equation}{0}
\section{The Schwinger operator formalism}

We performed the calculation in the Schwinger operator formalism\cite{Shuryak}. This
formalism is a very convenient method to do calculations involving covariant derivatives
because it preserves gauge invariance at all stages of the calculation. This is unlike
ordinary diagrammatic perturbation theory where the $\partial_{\mu}$ and the $g_s
G_{\mu}$ sitting in a covariant derivative appear at different orders in the expansion.
At the same time we shall regulate divergent integral with dimensional regularization by
dimensional reduction\cite{DR}. In this regularization one keeps the algebra of Dirac
matrices in four dimensions but the momentum integrals are kept D dimensional.

 One starts by defining quantum mechanical operators $\widehat{X}_{\mu},
\widehat{P}_{\nu}$ satisfying the eigenvalue equation
\begin{equation}\label{eigen}
    \widehat{X}_{\mu}|x\rangle=x_{\mu}|x\rangle\ ,
\end{equation}
and the commutation relations
\begin{eqnarray}\label{qm}
    \left[\widehat{X}_{\mu},\widehat{P}_{\nu}\right]&=&-i g_{\mu\nu}\nonumber\\
\left[\widehat{P}_{\mu},\widehat{P}_{\nu}\right]&=&ig_s\frac{\lambda^{a}}{2}G_{\mu\nu}^{a}\
,
\end{eqnarray}
where $\lambda^{a}$ are color Gell-Mann matrices satisfying
$\mathrm{Tr}\lambda^{a}\lambda^{b}=2 \delta^{ab}$ and $\widehat{P}_{\mu}=i
\partial_{\mu}+g_s\frac{\lambda^{a}}{2}G_{\mu}^a$ in position space. One has that
\begin{equation}\label{definition}
\langle x \vert {\widehat{P}\slash}\vert y\rangle=i{D}\slash_{\!\!\!x}\ \delta{(x-y)}\ ,
\end{equation}
where the covariant derivative is understood as acting upon the delta function. With this
definition the quark propagator can be expressed as:
\begin{equation}
S(k)=\int\,d^4x e^{ik\cdot x}\langle x\vert \frac{1}{\widehat{P}\slash-m_i}\vert
0\rangle=\int\, d^4x \langle x\vert \frac{1}{\widehat{P}\slash+k\slash-m_i}\vert 0\rangle
\end{equation}
where the second equality follows from the relation
\begin{equation}\label{relations}
    e^{ik\cdot\widehat{X}}\widehat{P}_{\nu}=(\widehat{P}_{\nu}+k_{\nu})
    e^{ik\cdot\widehat{X}}\ .
\end{equation}

With these definitions we can apply the formalism to $\Delta S=2$ transitions. Any quark
field $q(x)$ is to be understood as $q(\widehat{X})$ so that one has
$q(\widehat{X})|y\rangle=q(y)|y\rangle$. For example, let us look at the diagram depicted
in Fig.~3. The procedure can be straightforwardly extended to the remaining ones. The
contribution from this diagram can be obtained in coordinate space from the following
expression:
\begin{eqnarray}
&&\frac{G_F^{2}}{2}M_W^4 \lambda_u^2 \int d^4x\, d^4y\, d^4z\, d^4w\ \langle x\vert
\bar{s} (1+\gamma_5)\gamma^{\mu}\frac{1}{\widehat{P}\slash +k\slash -
  m_u}\gamma^{\nu}(1-\gamma_5)d \vert y\rangle \nonumber \\
  &&\langle x\vert \frac{1}{k^{2} - M_W^{2}}\vert z\rangle
\langle w\vert\bar{s}(1+\gamma_5)\gamma_{\nu}\frac{1}{\widehat{P}\slash
+k\slash-m_u}\gamma_{\mu}(1-\gamma_5)d\vert z\rangle\,\langle y\vert \frac{1}{k^{2} -
M_W^{2}}\vert w\rangle\ .
\end{eqnarray}

The integration of the W particle amounts to the expansion :
\begin{equation}
\langle x\vert \frac{1}{k^{2} - M_W^{2}}\vert z\rangle=\frac{-1}{M_W^{2}}\ \delta{(x-z)}+
...
\end{equation}
This has to be done twice. One of the Dirac deltas can be used to reduce the number of
integrals, whereas for the second one we use its representation in momentum space:
\begin{equation}
\delta{(x-y)}=\int \frac{d^4 k}{(2\pi)^4}e^{ik\cdot(x-y)}
\end{equation}
which will eventually give rise to the loop integral.

Now we can expand the quark propagators to any order in the soft momentum
$\widehat{P}\slash$ in the following way:
\begin{equation}
\frac{1}{\widehat{P}\slash + k\slash -m_u}=\frac{1}{k\slash
  -m_u}-\frac{1}{k\slash -m_u}\widehat{P}\slash \frac{1}{k\slash
  -m_u}+\frac{1}{k\slash -m_u}\widehat{P}\slash \frac{1}{k\slash -m_u}\widehat{P}\slash
\frac{1}{k\slash -m_u}+ ...
\end{equation}
Keeping terms of ${\cal{O}}(P^2)$, since we are interested in dimension-eight operators,
 and after some diracology, we end up with:
\begin{eqnarray}
&&\langle x \vert P_L \frac{1}{\widehat{P}\slash +k\slash - m_u}P_L\vert 0\rangle
=\langle x \vert P_L \frac{k\slash}{k^{2} -m_u^{2}}P_L\vert
0\rangle \nonumber\\
&&\qquad +\ \langle x \vert P_L \left[\frac{\widehat{P}\slash}{k^{2}
-m_u^{2}}-\frac{2(\widehat{P}\cdot k)k\slash}{(k^{2}
  -m_u^{2})^{2}}\right]P_L\vert 0\rangle \\
&&\!\!\!\!\!\!\!\!\!\!\!\!\!\!+\ \langle x \vert P_L \left[\frac{4 k\slash
(\widehat{P}\cdot
  k)^{2}}{(k^{2} -m_u^{2})^{3}}-\frac{\widehat{P}^{2} k\slash}{(k^{2} -m_u^{2})^{2}}
  -\frac{(\widehat{P}\cdot k)
  \widehat{P}\slash}{(k^{2} -m_u^{2})^{2}}-
  \frac{\widehat{P}\slash (\widehat{P}\cdot k)}{(k^{2} -m_u^{2})^{2}}-
  i\frac{\epsilon^{\alpha\beta\gamma\delta}
  \widehat{P}_{\alpha}\widehat{P}_{\beta}k_{\gamma}
  \gamma_{\delta}\gamma_5}{(k^{2} -m_u^{2})^{2}}
  \right]P_L \vert 0\rangle
  \nonumber
\end{eqnarray}
which is the expression to be employed for the $u$ quark propagator hereafter.

Now we use (\ref{definition}) to convert $\widehat{P}_{\mu}$ operators into covariant
derivatives. Integration by parts shifts the derivatives to the quark fields and delta
functions can then be integrated in a trivial way. Thus, one can write, for instance:
\begin{eqnarray}\label{oscar}
\int \frac{d^D k}{(2\pi)^D}\int d^4x\, d^4y\, \bar{s}_L(x)\gamma^{\mu}\langle x \vert
  \frac{k\slash}{k^2-m_u^2}\vert 0 \rangle
  \gamma^{\nu}d_L(0)\bar{s}_L(y)\gamma_{\nu}\langle
  y \vert \frac{\widehat{P}^2 k\slash}{(k^2-m_u^2)^2}\vert 0 \rangle
  \gamma_{\mu}d_L(0)=&&\nonumber\\
=\bar{s}_L(0)\overleftarrow{{D}^2}\gamma_{\nu}d_L(0)\bar{s}_L(0)\gamma_{\mu}d_L(0)\int
\frac{d^D
  k}{(2\pi)^D}\frac{k^{\mu}k^{\nu}}{(k^2-m_u^2)^3} &&\nonumber\\
=g_s\bar{s}_L(0){\tilde{G}}_{\nu\sigma}\gamma^{\sigma}d_L(0)
\bar{s}_L(0)\gamma_{\mu}d_L(0)\int \frac{d^D
  k}{(2\pi)^D}\frac{k^{\mu}k^{\nu}}{(k^2-m_u^2)^3} \qquad \qquad  &&
\end{eqnarray}
where the Dirac structure has already been simplified and the operator in the second line
has been rewritten in terms of the dual
\begin{equation}
{\tilde{G}}^{\mu\nu}=\frac{1}{2}\varepsilon^{\mu\nu\alpha\beta}G_{\alpha\beta}, \qquad
G_{\mu\nu}=  \frac{\lambda^a}{2}G_{\mu\nu}^a, \qquad \varepsilon^{0123}=1\ .
\end{equation}
In order to get to the result in Eq. (\ref{oscar}) one may use the identity
$\overleftarrow{D}^2 = \frac{g_s}{2}\ \sigma_{\mu\nu} G^{\mu\nu}+
{\overleftarrow{D}\slash}^2$, and the equations of motion for the quark fields in the
chiral limit. Furthermore, using
\begin{equation}\label{ident}
    \bar{s}_L{G}^{\alpha\nu}\gamma_{\nu}d_L=\frac{-i}{g}
    \left\{\bar{s}_L\left(D_{\alpha}D\slash- \overleftarrow{D}\slash D_{\alpha}\right)d_L -
    \partial_{\mu}\left( \bar{s}_L \gamma_{\nu} D_{\alpha}d_L\right) \right\}\ ,
\end{equation}
one finds that
\begin{equation}\label{ident2}
    \langle \overline{K}^0(p)|\overline{s}_L\gamma^\nu G_{\alpha\nu} d_L|0\rangle =
    \frac{i}{g}\ \partial_{\mu}\langle \overline{K}^0(p)|\overline{s}_L\gamma^\mu
    D_{\alpha}d_L|0\rangle \sim \mathcal{O}(p^2p_{\alpha})
\end{equation}
due to Eq. (\ref{current}) and, thus, it can be neglected.

Upon performing the integral over $k$ in Eq. (\ref{oscar}) one can easily extract the
$\mu$ dependence. After adding up all the contributions, one obtains the result
(\ref{eleven}) in the text.

\renewcommand{\theequation}{B.\arabic{equation}}
\setcounter{equation}{0}
\section{Ward Identities}

Let the Green's function $\Gamma_{\mathrm{\mathcal{A}}}^{\mu\nu\alpha}(q,p)$ be
\begin{equation}\label{gama}
\Gamma^{\mathrm{\mathcal{A}}}_{\mu\nu\alpha}(q,p) \equiv \int d^4x d^4y\ e^{-i(qx+py)}
    \langle 0|\mathrm{T}\left\{R_{\alpha}^{ds}(y)L_{\mu}^{su}(x)L_{\nu}^{ud}(0)
    \right\}|0\rangle\ ,
\end{equation}
where in this expression the symbol $\mathrm{T}\{...\}$ stands for the ordinary covariant
chronological product. Because of the chiral anomaly one has the following Ward
identities:
\begin{eqnarray}\label{cond1}
q_{\mu}\Gamma_{\mathrm{\mathcal{A}}}^{\mu\nu\alpha}(q,p)&=&-i\,\, \Pi^{\nu\alpha}_{LR}(p)
- i \frac{N_c}{48 \pi^2}\
\varepsilon^{\lambda\sigma\nu\alpha} q_{\lambda}p_{\sigma}\nonumber\\
(q+p)_{\nu}\Gamma_{\mathrm{\mathcal{A}}}^{\mu\nu\alpha}(q,p)&=&i\,\,
\Pi^{\mu\alpha}_{LR}(p) + i \frac{N_c}{48 \pi^2}\
\varepsilon^{\lambda\sigma\alpha\mu} q_{\lambda}p_{\sigma} \nonumber\\
p_{\alpha}\Gamma_{\mathrm{\mathcal{A}}}^{\mu\nu\alpha}(q,p)&=& i \frac{N_c}{24 \pi^2}\
\varepsilon^{\mu\nu\lambda\sigma} q_{\lambda}p_{\sigma}\ .
\end{eqnarray}

One can also define a similar Green's function to that in Eq. (\ref{gama}) but with a
different T product, which we shall denote by $\widehat{T}\{....\}$, which produces the
left-right, or factorized, form of the anomaly\cite{Bardeen,KPPdeR04}. In this case the
new Green's function, which we shall denote by $\Gamma_{\mu\nu\alpha}(q,p)$, satisfies
the naive Ward identities \emph{without} the anomaly terms. In other words, one has that
\begin{eqnarray}\label{cond2}
q_{\mu}\Gamma^{\mu\nu\alpha}(q,p)&=&-i\,\, \Pi^{\nu\alpha}_{LR}(p)\nonumber\\
(q+p)_{\nu}\Gamma^{\mu\nu\alpha}(q,p)&=&i\,\, \Pi^{\mu\alpha}_{LR}(p) \nonumber\\
p_{\alpha}\Gamma^{\mu\nu\alpha}(q,p)&=& 0\ .
\end{eqnarray}
The relationship between the two Green's functions is given by the equality
\begin{equation}\label{relat}
    \Gamma^{\mu\nu\alpha}(q,p)= \Gamma_{\mathcal{A}}^{\mu\nu\alpha}(q,p) +
     i \frac{N_c}{24 \pi^2}\
    \varepsilon^{\mu\nu\alpha\lambda} q_{\lambda} +
    i \frac{N_c}{48 \pi^2}\ \varepsilon^{\mu\nu\alpha\lambda} p_{\lambda}\ .
\end{equation}
As one can see in the equation above, they differ by a local counterterm, i.e. a
polynomial in momenta.

 For small momentum $p$, one can write down the most general
structure for $\Gamma^{\mu\nu\alpha}(q,p)$ compatible with the conditions in Eq.
(\ref{cond2}). Defining, as usual,
\begin{equation}
\Pi^{\nu\alpha}_{LR}(p)\equiv \int\,d^4x\,e^{-ip\cdot y}\langle 0 \vert
T\left\{R_{ds}^{\alpha}(y)L_{sd}^{\nu}(0)\right\}\vert 0
\rangle=\left(g^{\nu\alpha}-\frac{p^{\nu}p^{\alpha}}{p^2}\right)\Pi_{LR}(p^2)\ ,
\end{equation}
and recalling that
\begin{equation}\label{pifunction}
    \Pi_{LR}(0)= -i\ \frac{F_0^2}{2}\ ,
\end{equation}
we find the following form for $\Gamma^{\mu\nu\alpha}(q,p)$ for small $p$:
\begin{equation}
  \Gamma_{\mu\nu\alpha}(q,p) = \Pi_{LR}(0){\cal{T}}_{\mu\nu\alpha}^{S}
   + I_1(Q^2){\cal{T}}_{\mu\nu\alpha}^{ST}
   + I_2(Q^2){\cal{T}}_{\mu\nu\alpha}^A + \mathcal{O}(p)\ ,
\end{equation}
where $Q^2=-q^2$, and
\begin{eqnarray}\label{tensors}
{\cal{T}}_{\mu\nu\alpha}^{S}&=&\frac{i}{p^2 q^2}
  \left[p^2 q_{\mu} g_{\alpha\nu}+p^2 q_{\nu}g_{\alpha\mu}-p_{\alpha}p_{\nu}q_{\mu}
  -p_{\alpha} p_{\mu} q_{\nu}+ q_{\mu} q_{\nu}
  \left(\frac{p\cdot q }{q^2}p_{\alpha}-\frac{p^2}{q^2}q_{\alpha}\right)\right]\nonumber\\
{\cal{T}}_{\mu\nu\alpha}^{ST}&=&\left(q^2 g_{\mu\nu} -q_{\mu}q_{\nu}\right)
   \left(\frac{p\cdot q}{p^2}p_{\alpha}-q_{\alpha}\right)\nonumber\\
{\cal{T}}_{\mu\nu\alpha}^A&=&\left[i \varepsilon_{\mu\nu\lambda\sigma}
   q^{\sigma} \frac{p_{\alpha}p^{\lambda}}{p^2}-
   i  \varepsilon_{\mu\nu\alpha\lambda} q^{\lambda}  \right]\ .
\end{eqnarray}
This is the expression (\ref{nineteen}) in the main text. In Eq. (\ref{tensors}), the
superindices $S,T,A$ summarize the symmetry properties of each tensor with respect to the
exchange of the indices $\mu,\nu$: ${\cal{T}}_{\mu\nu\alpha}^{S}$ is symmetric,
${\cal{T}}_{\mu\nu\alpha}^{ST}$ is also symmetric and, furthermore, transverse with
respect to $q_{\mu}$. Finally, ${\cal{T}}_{\mu\nu\alpha}^{A}$ is $\mu\nu$ antisymmetric.

At low momentum one finds that
\begin{equation}\label{low}
    I_1(Q^2)= 2 \frac{L_9}{Q^2}+ \mathcal{O}(Q^0)\quad ,
    \quad I_2(Q^2)= - \frac{N_c}{24 \pi^2}+ \mathcal{O}(Q^2)\ ,
\end{equation}
whereas, at large momentum one has
\begin{equation}\label{high}
    I_1(Q^2)= \frac{F_0^2}{2Q^4}+ \mathcal{O}(\frac{1}{Q^6})\quad ,
    \quad I_2(Q^2)= - \frac{F_0^2}{2 Q^2}+ \mathcal{O}(\frac{1}{Q^4})\ .
\end{equation}

As a consequence of the symmetry properties of the tensors $\mathcal{T}_{\mu\nu\alpha}$
there are no crossed terms in the following contraction:
\begin{eqnarray}
&&\int d\Omega_{q}\ \Gamma^{\mu\nu\alpha}(q,p)\Gamma_{\nu\mu\beta}(q,p)=
\\
&&\left(\frac{p^{\alpha}p_{\beta}}{p^2}-g^{\alpha}_ {\beta}\right) \left\{\frac{3}{4}\
I_1^2(Q^2) \ Q^6- \frac{3}{2} \ I_2^2(Q^2)\ Q^2
    + \frac{7}{16}\ \frac{F_0^4}{Q^2}\right\}
    + \mathcal{O}(p^2)\nonumber \ ,
\end{eqnarray}
and Eq. (\ref{twenty}) in the main text follows.

\end{document}